\documentclass[aps,prl, reprint,amsmath,amssymb, superscriptaddress]{revtex4-1}
\usepackage{graphicx} 
\usepackage{amssymb,amsfonts,amsmath}
\usepackage{txfonts}
\usepackage{xcolor}              
\usepackage{setspace}            
\usepackage{fullpage}            
\usepackage{url}
\usepackage[caption=false]{subfig}
\usepackage{booktabs}
\usepackage{url}
\usepackage{placeins}
\usepackage{epstopdf}
\usepackage{ragged2e}
\usepackage{lipsum}

\definecolor{darkgreen}{HTML}{00BB00}

\def\er{Erd\H{o}s-R\'enyi }

\begin{document}
\title{Percolation of Hierarchical Networks and Networks of Networks}
\date{\today}
\author{Louis M. Shekhtman}
\affiliation{ Department of Physics, Bar-Ilan University, Ramat Gan, Israel}
\author{Shlomo Havlin}
\affiliation{ Department of Physics, Bar-Ilan University, Ramat Gan, Israel}

\begin{abstract} 
   Much work has been devoted to studying percolation of networks and interdependent networks under varying levels of failures. Researchers have considered many different realistic network structures, but thus far no study has incorporated the hierarchical structure of many networks. For example, infrastructure across cities will likely be distributed such that nodes are tightly connected within small neighborhoods, somewhat less connected across the whole city, and have even fewer connections between cities. Furthermore, while previous work identified interconnected nodes, those nodes with links outside their neighborhood, to be more likely to be attacked, here we have various levels of interconnections (between neighborhoods, between cities, etc.). We consider the nodes with interconnections at the highest level most likely to be attacked, followed by those with interconnections at the next level, etc. We develop an analytic solution for both single and interdependent networks of this structure and verify our theory through simulations. We find that depending on the number of levels in the hierarchy there may be multiple transitions in the giant component (fraction of interconnected nodes), as the network separates at the various levels. Our results show that these multiple jumps are a feature of hierarchical networks and can affect the vulnerability of infrastructure networks.
   
\end{abstract}
\maketitle
\section{Introduction}
The robustness of infrastructure systems can be understood through the frameworks of complex networks, percolation, and interdependent networks \cite{albert2000error,cohen-prl2000,albert-rmp2002,newman-book2009,gao-naturephysics2012,baxter-prl2012,kivela2014multilayer,de2013mathematical,de2016physics,boccaletti2014structure}. The initial research on network robustness was later expanded to include various network structures such as various degree distributions \cite{zhou2013percolation,cohen2003structural,goltsev2008percolation}, clustering \cite{newman2003properties,shao2014robustness}, spatial embedding \cite{bashan-naturephysics2013,wei-prl2012,shekhtman2014robustness}, and quite recently, community structure \cite{shai2015critical,shekhtman2015resilience}. Further, additional research has considered various types of attacks on these networks such as degree-based attacks \cite{cohen-prl2001,callaway-prl2000}, localized attacks \cite{berezin2015localized,shao2015percolation}, and attacks based on nodes linking across communities \cite{shai2015critical,shekhtman2015resilience,da2015fast}.

However, other common network structures that are likely relevant for robustness have not yet been studied. Among these is a hierarchical structure, which we will study here, where communities connect loosely with one another to form larger communities and so on \cite{lancichinetti2009detecting,clauset2008hierarchical,palla2005uncovering} (See Fig. \ref{fig:model}). In the context of infrastructure robustness these overlapping modules are likely described through neighborhoods overlapping to form cities, which then overlap to form states, etc. which are then interconnected among themselves. 

Furthermore, in this model, the nodes at the highest level of the hierarchy (e.g. between states) are likely more vulnerable to failure or attack than those at the next highest level, which are in turn more vulnerable than those at an even lower level etc. This is because the nodes at higher levels have longer distance links between them which are more likely to fail or be attacked \cite{mcandrew2015robustness} and also have higher betweenness \cite{shai2015critical} which yields additional load on them \cite{zhao2016spatio,motter2002cascade}. Also, recent work by da Cunha et al. \cite{da2015fast} showed that attacks on these  `interconnected nodes' are an optimal form of attack on the US power grid, an infrastructure system of critical interest.

\section{Model}
Our model is a stochastic block model \cite{peixoto2012entropy,peixoto2014hierarchical,decelle2011asymptotic,karrer2011stochastic} with overlap among the various blocks. 

\begin{figure}
	\centering
	\includegraphics[width=0.8\linewidth]{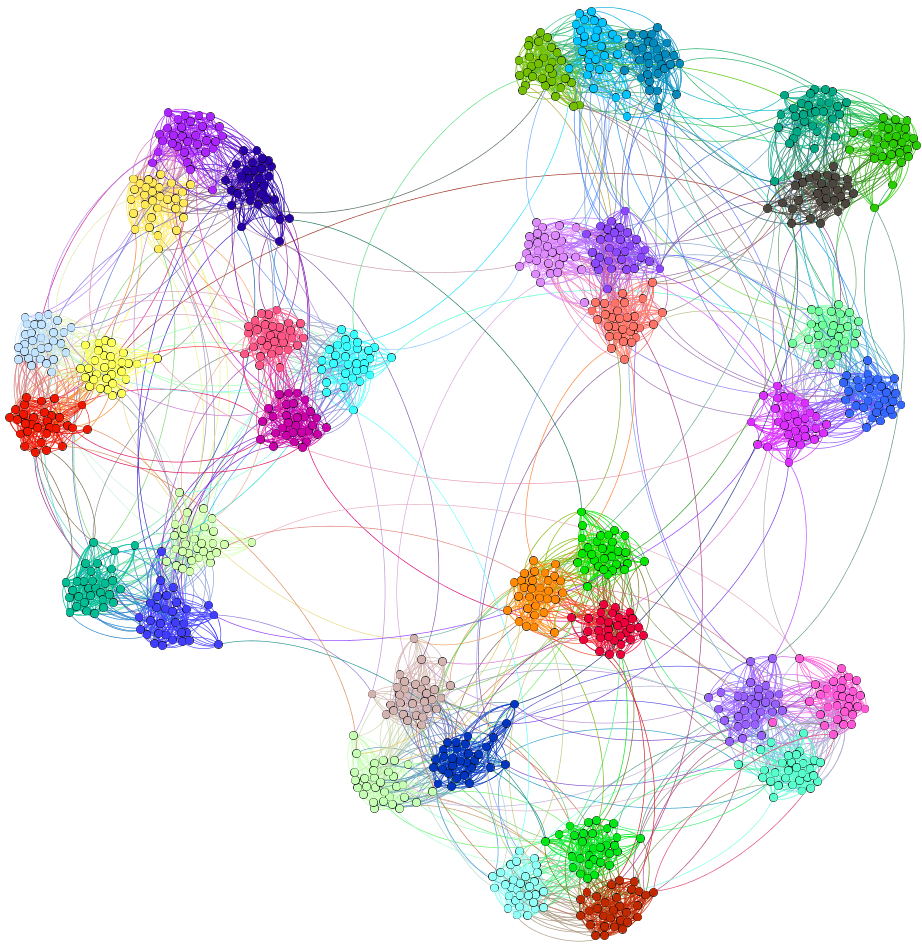} 
	\caption{\textbf{Model Illustration.} For this realization, the model has 4 hierarchical layers. At the top layer there are 3 modules, each of which is broken down into 4 modules, each of which is then broken down into 3 modules, which are not broken down further. We could describe this configuration of modules by the vector $\vec{m}=\left[1,3,12,36\right]$ }
	\label{fig:model}
\end{figure}
We first define the vector $\vec{m}$ describing the number of distinct modules or communities (blocks) at each layer. At the first layer we always consider the entire network as a single community, thus $m_1=1$. The next layer, $m_2$ counts how many modules are at the top layer. Next is the total number of modules at the third layer, next is the modules at the fourth layer, etc (we also assume for simplicity that all of the $m_j$ modules are broken down into the same fixed number of $m_{j+1}$ modules). For example, if we take the network shown in Fig. \ref{fig:model}, we would say $\vec{m}=\left[1,3,12,36\right]$, since the top layer is a connected graph, at the next layer we have three modules, then a total of 12 modules (i.e. each of the three is broken down into four smaller modules and $3\times4=12$), and finally 36, since each of the 12 modules is broken down into three additional ones. 

We next define the vector $\vec{k}$, which describes the average degree between nodes connected at each layer of the network. Thus, if at the highest layer there is an average of $0.1$ links between each module, this will be the first entry, $k_1$ in $\vec{k}$. If the average degree at the next layer is $0.3$ then that will be the second entry, $k_2$ etc. We assume that the entries of $\vec{k}$ should be strictly increasing since we expect there to be more links within communities at a lower layer than at a higher layer (e.g. neighborhoods are more tightly connected than cities).

We will carry out a targeted attack on the nodes of the network, assuming that nodes that are interconnected at the top level are most likely to fail. To do so, we must determine how many nodes are connected at each level and convert from their survival likelihood to the overall survival likelihood. We can estimate how many nodes are connected at level $i$ as $1-e^{-k_i}$ \cite{shai2015critical}. We then define $r_i$ as the survival probability of interconnected nodes at layer $i$. For the top layer of interconnections we can convert from $r_1$, the survival probability of interconnected nodes at the highest level, to $p$, the overall survival probability, using \cite{shai2015critical}
\begin{equation}
	r_1=\frac{p-e^{-k_1}}{1-e^{-k_1}}.
	\label{eq:r-value}
\end{equation}

After we have removed all nodes with interconnections at the top level, we then begin removing those nodes with interconnections at the next level. In order to convert from the survival probability of nodes at this next level, $r_2$ and the new overall survival probability $p$, we must take into account those nodes removed at the previous layer. 

We do so by first finding the value of $p$ for which we have removed all nodes at a given layer. For example, for the first layer the cutoff for which all nodes with interconnections at this layer are removed is $p_{co_1}=e^{-k_1}$. For the next layer the cutoff is given by recognizing that the number of nodes with interconnections at this next layer is $e^{-k_2}$, but we are already at a survival probability of only $p_{co_1}$ and also some of these nodes also had interconnections at the previous layer. Thus taking these into account gives that the cutoff of $p$ at the second layer, $p_{co_2}$ is given by the inclusion-exclusion principle as
\begin{align}
p_{co_2}&=1-\left(1-e^{-k_2}+1-e^{-k_1}-(1-e^{-k_2})(1-e^{-k_1})\right) \nonumber \\
&=e^{-k_2-k_1}\label{eq:cutoff2}.
\end{align}
Thus the cutoff value of $p$, for a given level $i$ is given by
\begin{equation}
p_{co_i}=e^{-\sum_{j=1}^i k_j}.
\label{eq:cutoff-i}
\end{equation}

We can then convert from $r_i$, the survival probability in level $i$ (after having removed all nodes in higher layers), to $p$ using
\begin{equation}
	r_i=\frac{p-p_{co_i}}{p_{co_{i-1}}-p_{co_i}}.
	\label{eq:r-level-i}
\end{equation}

\begin{figure}[htbp]
	\centering
	\subfloat{%
		\includegraphics[width=0.47\textwidth]{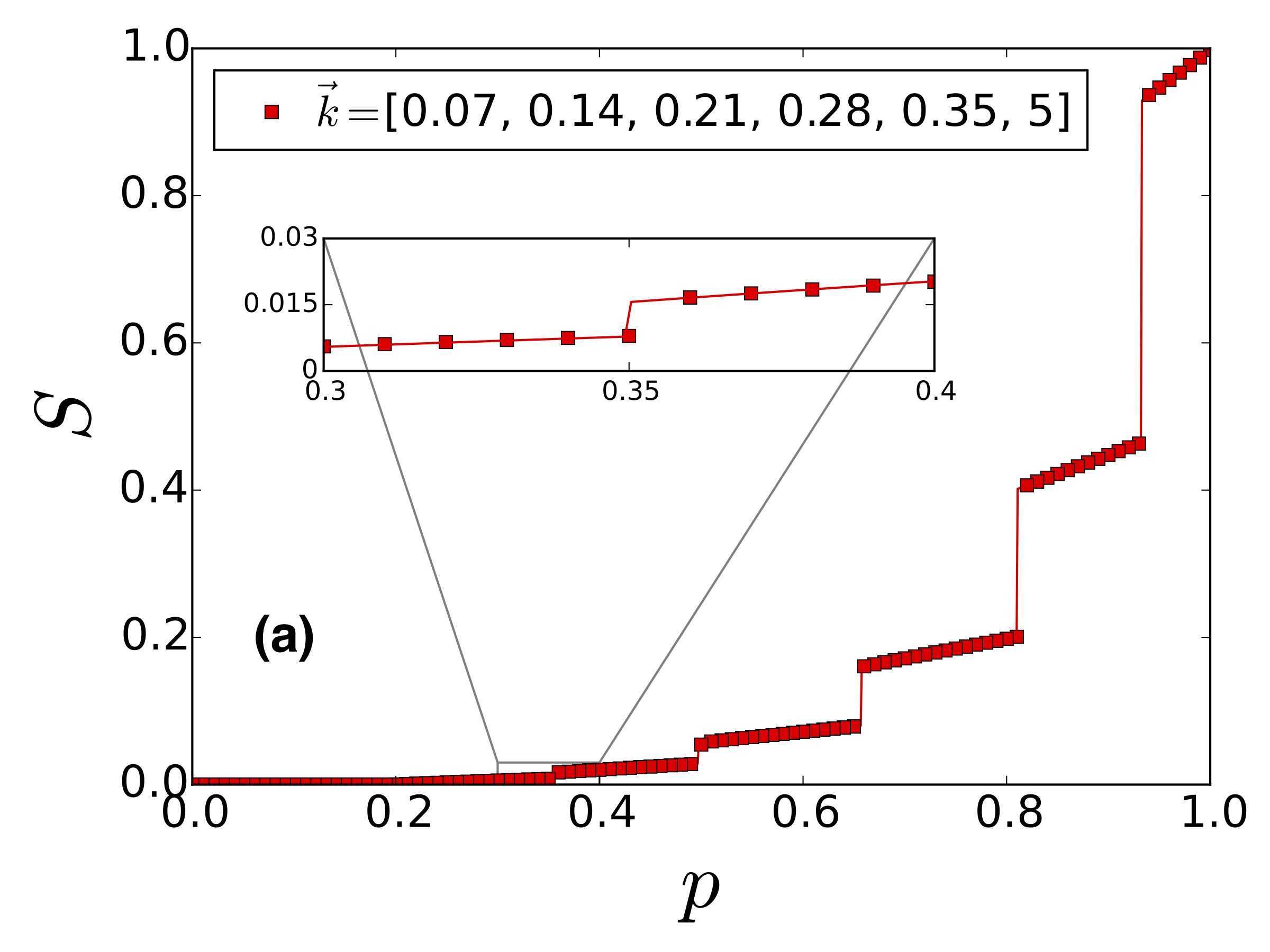} 		\label{fig:single-layer1}
	}\hfill
	\subfloat{
		\includegraphics[width=0.47\textwidth]{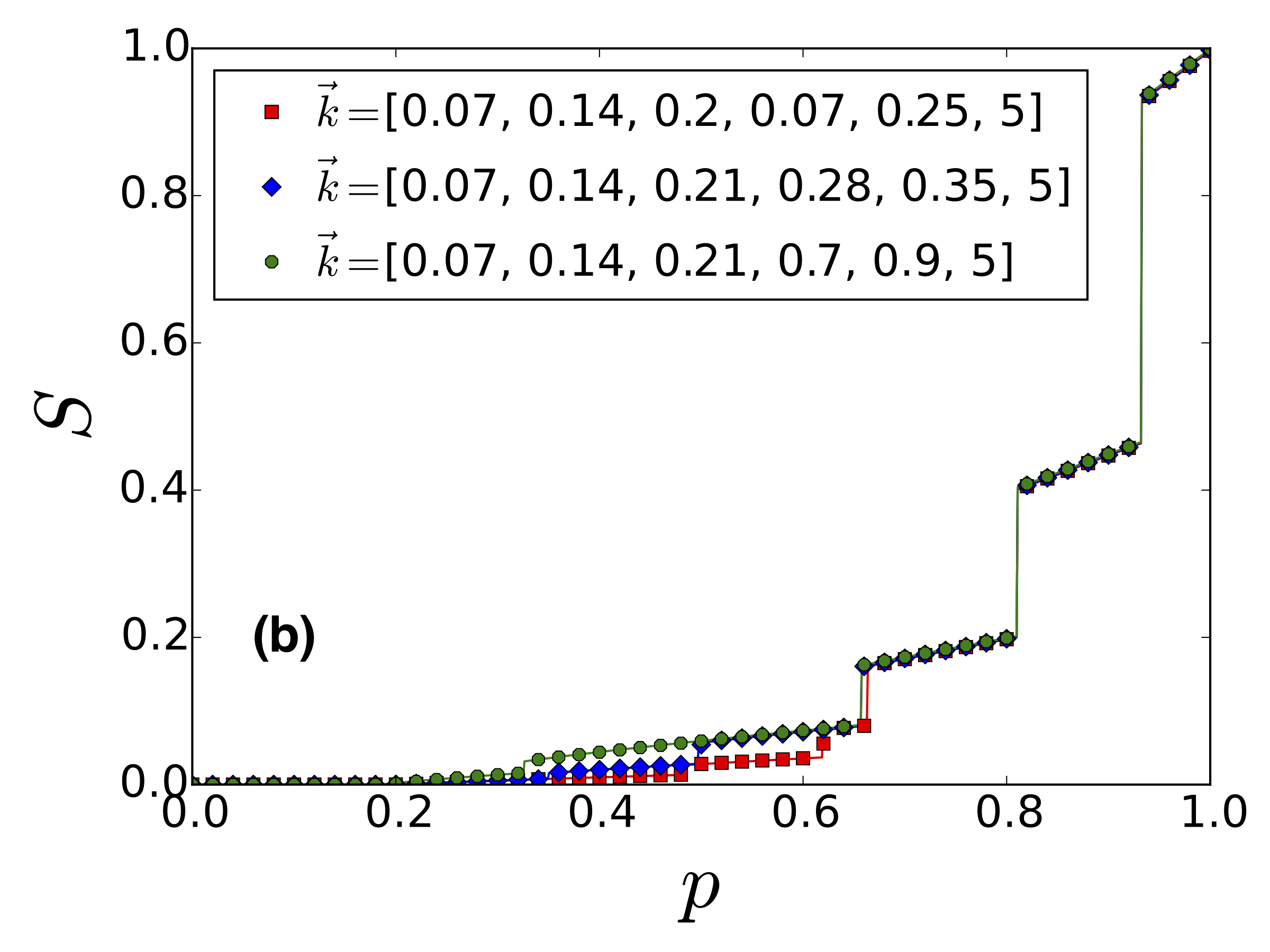} 
		\label{fig:single-layer2}}
	\caption{\textbf{Comparison of simulation results with theory.}  In both figures we have six layers, where each layer splits a module into two other modules, thus there are $\vec{m}=\left[1, 2, 4, 8, 16, 32\right]$ modules at each layer, with the average degrees between nodes at each layer given by the vector $\vec{k}$ in the legend. We have different values for the degree at each layer as given in the legends of \textbf{(a)} and \textbf{(b)}. The lines represent the theory outlined above and the points are simulations averaged over 10 runs on networks of $N=10^6$ nodes. }
	\label{fig:results}
\end{figure}

\section{Analytic Theory for a Single Network}
Having made these conversions, we can now find the size of the giant component after some fraction of nodes are removed. For the top level, we can extend previous results on modular networks by setting our average interconnected degree to the degree at the level we are attacking and setting our average intra-degree to the sum of the degrees at the lower levels \cite{shai2015critical,shekhtman2015resilience}. We thus obtain
\begin{multline}
P_\infty=e^{-k_1}(1-r_1)\left(1-e^{-\left(\sum_{i=2}^l k_i\right) P_\infty}\right)\\+r_1\left(1-e^{-\left(\sum_{i=1}^l k_i\right) P_\infty}\right),  \qquad {p>p_{{co}_0}}.
\label{eq:single-first-layer}
\end{multline}

Once we have removed all interconnected nodes in the first level, we then move on to removing nodes that are interconnected at the second level. In this case, the average degree of interconnections is now $k_2$ and the average degree of intraconnections is $\sum_{i=3}^l k_i$, where $l$ is the number of levels. Furthermore, for this case we will have removed some fraction $1-r_2$ of nodes based on targetted attack of interconnected nodes at this layer, and also $p_{co_1}$ nodes randomly due to the attacks at the above layer. Additionally, we must recall that the network at this stage is already split into $m_2$ separate modules thus we must scale $P_\infty$ by $1/m_2$. This gives
\begin{multline}
m_2P_\infty=p_{{co}_1}\Bigg[e^{-k_2}(1-r_2)\left(1-e^{-\left(\sum_{i=3}^l k_i\right) m_2 P_\infty}\right)\\+r_2\left(1-e^{-\left(\sum_{i=2}^l k_i\right) m_2 P_\infty}\right)\Bigg],  \qquad p_{{co}_1}>p>p_{{co}_2}.
\end{multline}

In general for all values of $p$ we can find the size of the largest connected component, $P_\infty$ using

\begin{multline}
m_jP_\infty=p_{{co}_{j-1}}\Bigg[e^{-k_j}(1-r_j)\left(1-e^{-\left(\sum_{i=j+1}^l k_i\right) m_j P_\infty}\right)\\+r_j\left(1-e^{-\left(\sum_{i=j}^l k_i\right) m_j P_\infty}\right)\Bigg],  \qquad p_{{co}_{j-1}}>p>p_{{co}_j}.
\label{eq:pinf-single-layer}
\end{multline}

We compare theory and simulations in Fig. \ref{fig:results}, observing excellent agreement between them. Also, in the figure we observe multiple discontinuities \cite{bianconi2014multiple,wu2014multiple} as once all interconnected nodes in a particular layer are removed, the system experiences a discontinuous jump.
 
\subsection{Number of Abrupt Jumps}
We next compute the expected number of jumps that will take place using our analytic theory from above. We begin by noting that jumps will occur when all interconnected nodes at a particular layer are removed, yet there remain enough total surviving nodes such that the hierarchy of modules at the next layer remains connected. This condition can be expressed by recognizing that we need the value of $p$ of the remaining intralinks to be lower than the value of $p$ for which all interconnected nodes at this given level are removed. This condition is given by for a given level $i$
\begin{equation}
e^{-k_i} \geq \frac{1}{\sum_{j=i+1}^l k_j}.
\label{eq:jump-condition}
\end{equation}
Assuming that the degree at each level of hierarchy is strictly decreasing, then $e^{-k_i}$ is strictly decreasing. Furthermore, assuming all $k_i>0$ implies that $1/\sum_{j=i+1}^l k_j$ is strictly increasing as $i$ increases (since the denominator must decrease as there are fewer $k_j$ terms). Therefore, once the condition of Eq.~\eqref{eq:jump-condition} is first violated for a particular level, we know that it will continue to break down for later levels and thus we can be sure that our number of jumps is the number of levels for which Eq.~\eqref{eq:jump-condition} is valid.

We plot the two sides of Eq.~\eqref{eq:jump-condition} in Fig. \ref{fig:single-jumps}, where for $l\leq 5$ we see that the left-hand-side (LHS) of the equation is larger than the right-hand-side (RHS). Comparing to the number of jumps in Fig. \ref{fig:single-layer1} we see that the network indeed experiences 5 abrupt jumps as expected (see inset for the 5th jump).

\subsection{The $p$ Values of the Jumps}
Having found the number of jumps that the network will undergo above, we can now analyze the multiple values of $p_c$, the critical thresholds at which the jumps occur. We first note that so long as the LHS of Eq.~\eqref{eq:jump-condition} is greater than the RHS of the same equation, then there will be a transition at the point of the LHS of the equation. After these $i$ transitions, there will be one final $i+1$st transition \cite{shekhtman2015resilience} at the point $r_{i+1}$, which can be found by solving the below equation for $r_{i+1}$,
\begin{widetext}
\begin{multline}
r_{i+1}^2 \left(  p_{co_i}^2 \left(\sum_{j=i+2}^l k_j\right) k_{i+1} e^{-k_{i+1}}   \right) +  r_{i+1}  \left(  p_{co_i} k_{i+1} + p_{co_i} \left(\sum_{j=i+2}^l k_j\right) - p_{co_i} \left(\sum_{j=i+2}^l k_j\right) e^{-k_{i+1}} - p_{co_i}^2 \left(\sum_{j=i+2}^l k_j\right) k_{i+1} e^{-k_{i+1}} \right) \\+ \left( p_{co_i} \left(\sum_{j=i+2}^l k_j\right) e^{-k_{i+1}} -1 \right) = 0.
\end{multline}
\end{widetext}

\begin{figure}
	\centering
	\subfloat{
		\includegraphics[width=0.9\linewidth]{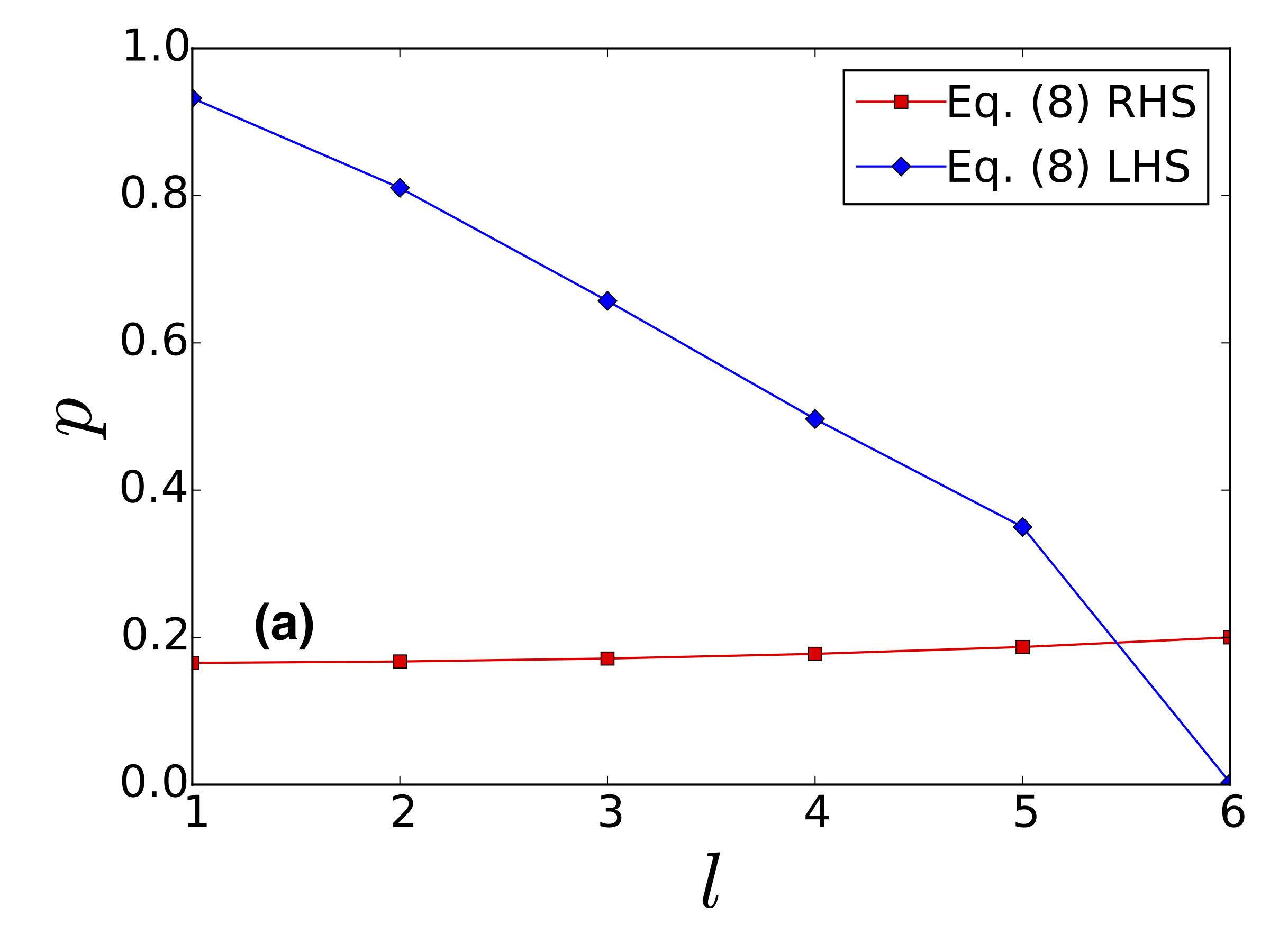} 	\label{fig:single-jumps}}
	\vfill
	\subfloat{
		\includegraphics[width=0.9\linewidth]{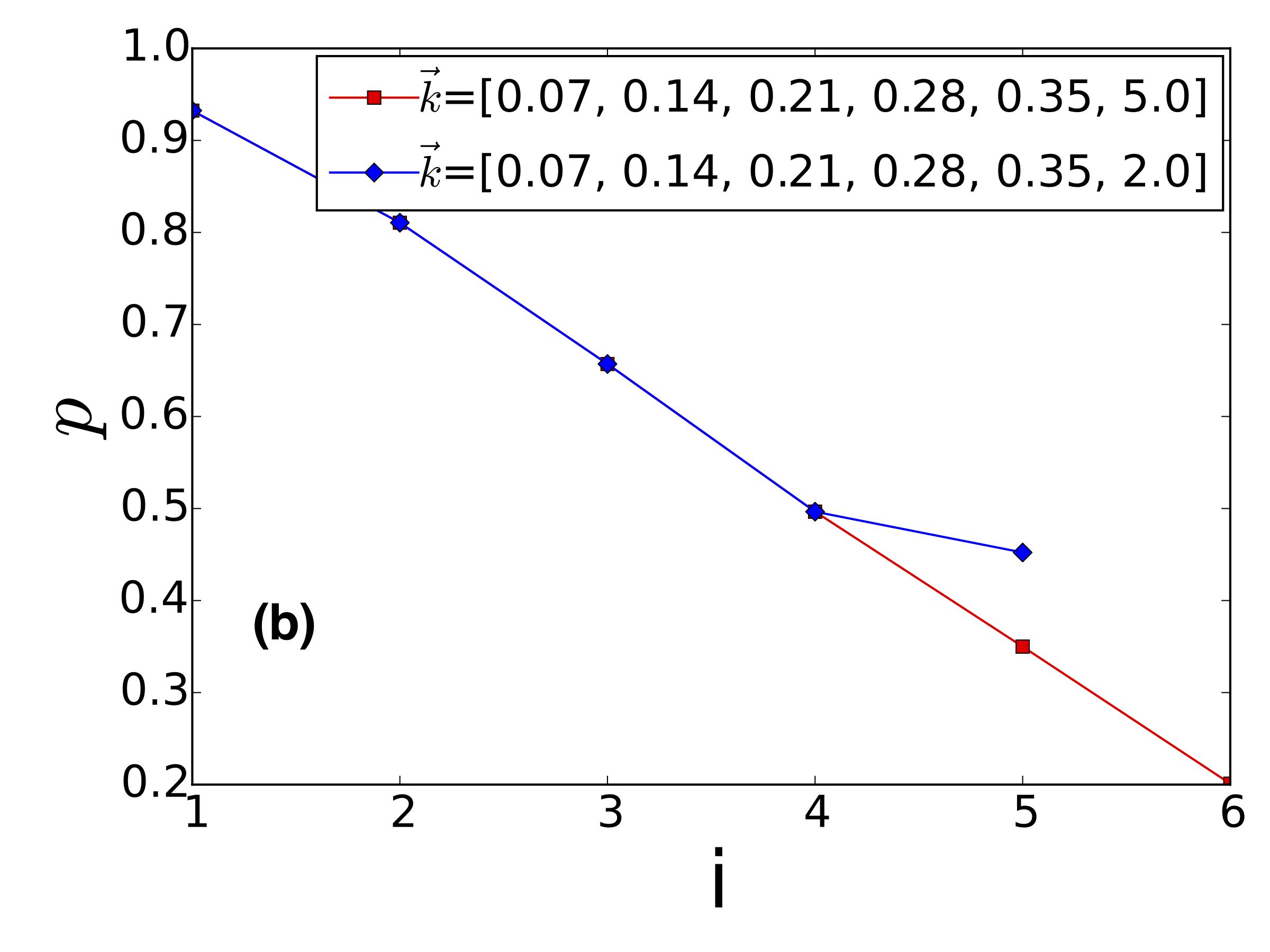} 	\label{fig:pc-vals}}
	
	\caption{\textbf{(a)} The RHS and LHS of Eq.~\eqref{eq:jump-condition} for the network of Fig.~\ref{fig:single-layer1}. We observe that for the first $l=5$ layers the value of $p$ for which all interconnected nodes are removed (LHS of Eq.~\eqref{eq:jump-condition}) is greater than the value at which the network connectivity as a whole breaks down (RHS of Eq.~\eqref{eq:jump-condition}). However for the 6th layer this is no longer true and we observe the continuous percolation transition of a random network. \textbf{(b)} The values of the $i$th critical points for both the network described in Fig.~\ref{fig:single-layer1} and another network which has a smaller degree at its bottom level. We note how changing the degree at the bottom level effects the number of jumps since for the network with a lowest level average degree of $2$, before removing all interconnected nodes at the 5th layer, the network already breaks apart.} 
	\label{fig:jumps-pc}
	
\end{figure}

After finding $r_{i+1}$ we can convert it to a value of $p$ using Eq.~\eqref{eq:r-level-i}. We note a slight subtlety in this system, in that even for the case where the hierarchical network is completely isolated at the lowest level we do not precisely recover percolation on a random network since we are  targeting only those nodes which began with at least one link in the construction of the network. This leads to a slight correction where we obtain $r_{i+1}=1/k_{i+1}$ (and then convert this to a value for $p_c$), rather than obtaining the usual $p_c=1/k_{i+1}$. In most cases this correction will be quite small as for any reasonable value of $k$ at the lowest level, there will be very few nodes that do not have a single link.  For example, for the case of the network referred to by the top line of the legend in Fig. \ref{fig:pc-vals}, the transition for the $6$th layer takes place at $p_c\approx0.201$ as opposed to $1/k_6=0.2$. 

\section{Interdependent Networks}
Much recent research has also explored the resilience of interdependent networks where the nodes of one network depend on nodes in another network \cite{buldyrev-nature2010,gao-naturephysics2012,bianconi2013statistical,bianconi2016percolation,brummitt2012multiplexity,van2016interconnectivity,gao-prl2011,min2014network,cellai-pre2013}. One example is that of a communication network that is interdependent with a power grid, yet more complex interdependencies are also possible \cite{rosato-criticalinf2008,rinaldi-ieee2001}. Many of these interdependent networks will likely possess the hierarchical structure described above. Therefore we now extend our theory to the case of networks of interdependent networks (NON). 

We will assume that each network is formed of the same hierarchical structure, i.e. there are the same number of modules at each level. Again, this is intuitive since the number of cities, neighborhoods, etc. that exist for the power grid are likely the same as those for a communications network. Further, we will assume that nodes are dependent on other nodes within their same module at the lowest level. This corresponds to the assumption that nodes are most likely dependent for resources from nodes in their same neighborhood, i.e. a power station depends on a communication tower in the same neighborhood.

In the case of interdependent networks formed of $n$ networks with $n>2$, the structure of the dependencies can take various shapes. Among these are both treelike structures, where networks depend on one another such that their dependencies form a tree, or looplike structures where the dependencies form loops. Here we will consider (i) treelike structures and (ii) a random-regular (RR) network of networks where each network depends on exactly $w$ other networks.

\subsection{Treelike Network Formed of Hierarchical Networks}

Here we introduce the theory for a network of networks formed of $n$ interdependent networks such that they form a tree. For the interdependent case, Eqs.~\eqref{eq:r-level-i}-\eqref{eq:single-first-layer} remain valid as for the single-network case. 

For the treelike NON we assume that all nodes between pairs of interdependent networks have a dependent node. Further, we assume the no-feedback condition, meaning that if node $a$ in network $n_1$ depends on node $b$ in network $n_2$, then node $b$ also depends on node $a$. Lastly, we will attack the nodes of only one of the networks and let the attack propagate to the other networks.

\begin{figure}[htbp]
	\centering
	
	\subfloat{
		\includegraphics[ width=1.0\linewidth]{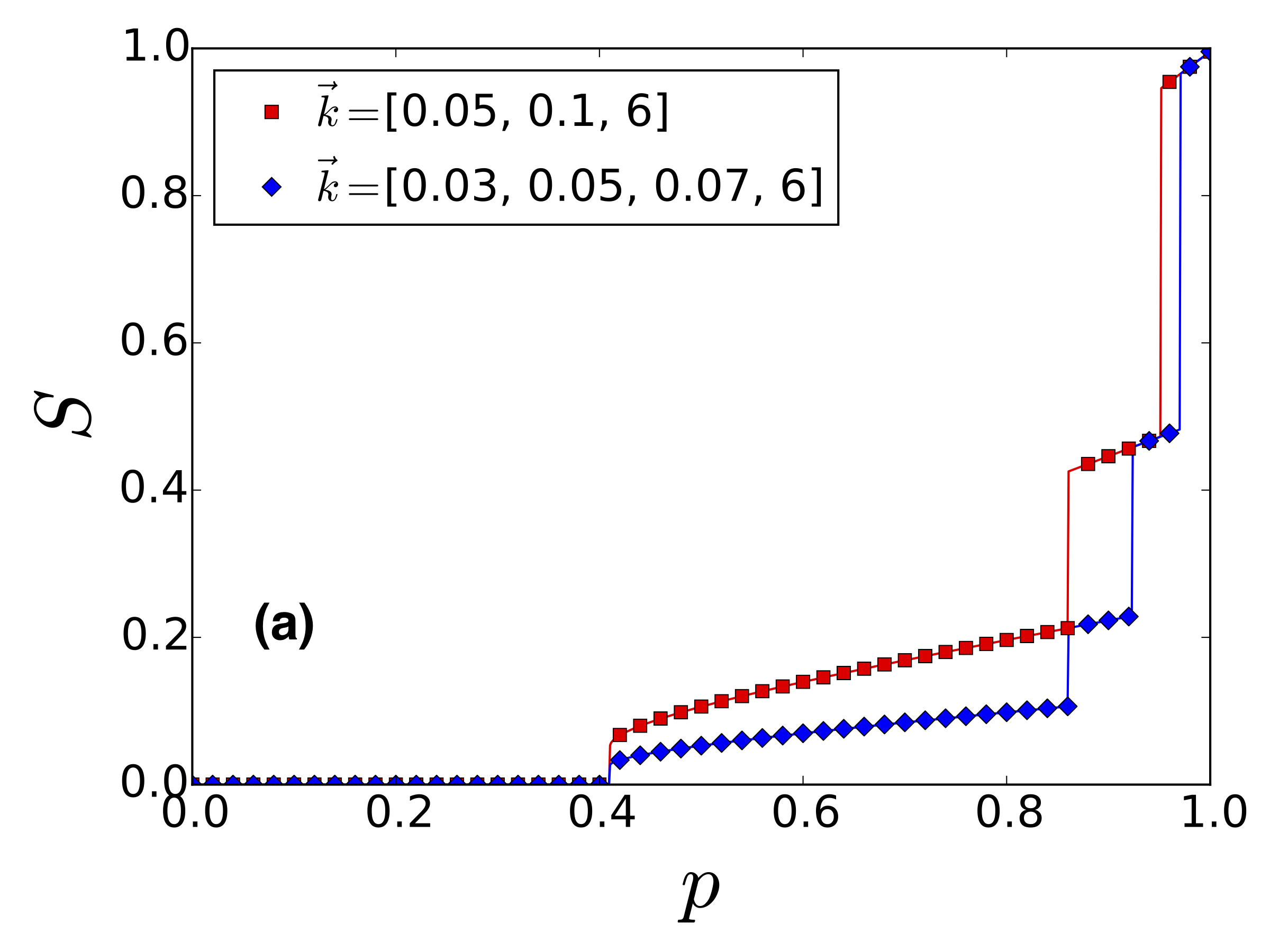}  \label{fig:tree-varyk} 
	}

	\hfill
	\subfloat{
		\centering
		\includegraphics[width=1.0\linewidth]{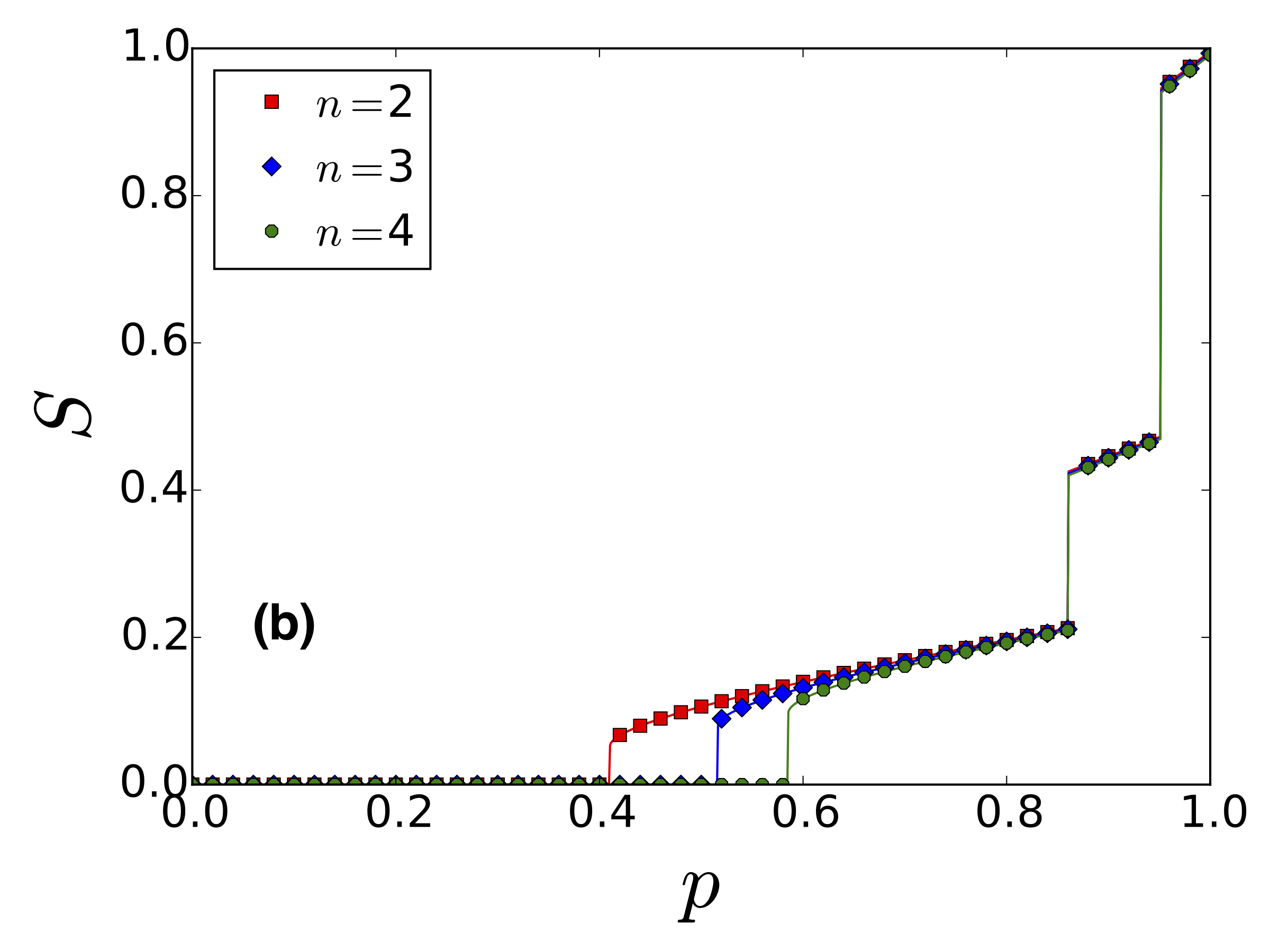} 
		\label{fig:tree-varyn}
	}
	\caption{\textbf{(a).} The case of two interdependent networks with the values of degree at each layer as given in the legend. Points are simulations averaged over 10 realizations of networks with $N=10^6$ nodes and lines are theory. \textbf{(b).} Varying the number of networks with the degree vector fixed to $\vec{k}=(0.05,0.1,6)$.}
	\label{fig:tree}
\end{figure}

To include the effects of the interdependencies, we note that we must add an additional likelihood of failure based on the interdependence. For a treelike network of $n$ interdependent networks with the dependencies within the neighborhoods, this term is $\left(1-e^{-\left(\sum_{i=j}^l k_i\right)m_jP_\infty}\right)^n$ \cite{gao-prl2011,shekhtman2015resilience}. We then multiply this term with the other terms of Eq.~\eqref{eq:pinf-single-layer} to obtain

\begin{multline}
	m_jP_\infty=p_{{co}_{j-1}}\Bigg[e^{-k_j}(1-r_j)\left(1-e^{-\left(\sum_{i=j+1}^l k_i\right) m_j P_\infty}\right)\\+r_j\left(1-e^{-\left(\sum_{i=j}^l k_i\right) m_j P_\infty}\right)\Bigg]\left(1-e^{-\left(\sum_{i=j}^l k_i\right)m_jP_\infty}\right)^{n-1}, \\ \qquad p_{{co}_{j-1}}>p>p_{{co}_j}.
\end{multline}
We note that numerical simulations show excellent agreement with the theory (Fig.~\ref{fig:tree}). We also note that in the case of interdependent networks, the final transition is now discontinuous due to the interdependence \cite{gao-naturephysics2012,gao-prl2011,buldyrev-nature2010}.

\subsection{Random Regular Network Formed of Hierarchical Networks}
Lastly, we consider the case of an RR NON where each network depends on exactly $w$ other networks. We will assume that for each pair of interdependent networks only a fraction $q$ of the nodes are interdependent and we will allow  feedback (in contrast to what was done for treelike NON). However, we will still restrict the dependencies such that they must be within the same community at the lowest level of the hierarchy.

\begin{figure}
	\centering
	\subfloat{
	\includegraphics[width=0.9\linewidth]{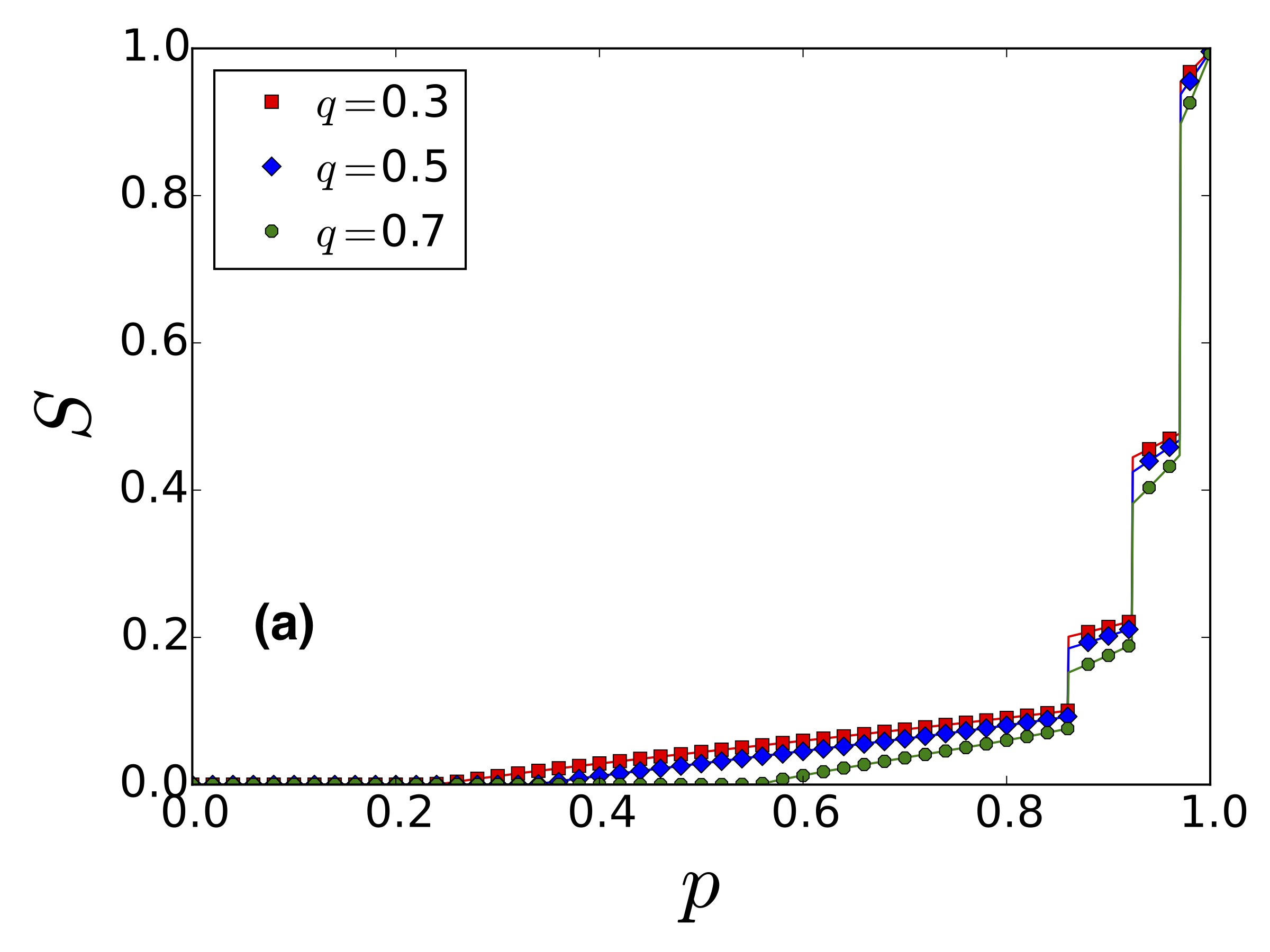} 
		\label{fig:RR_varyq}
	}
	\vfill
	\subfloat{
		\includegraphics[width=0.9\linewidth]{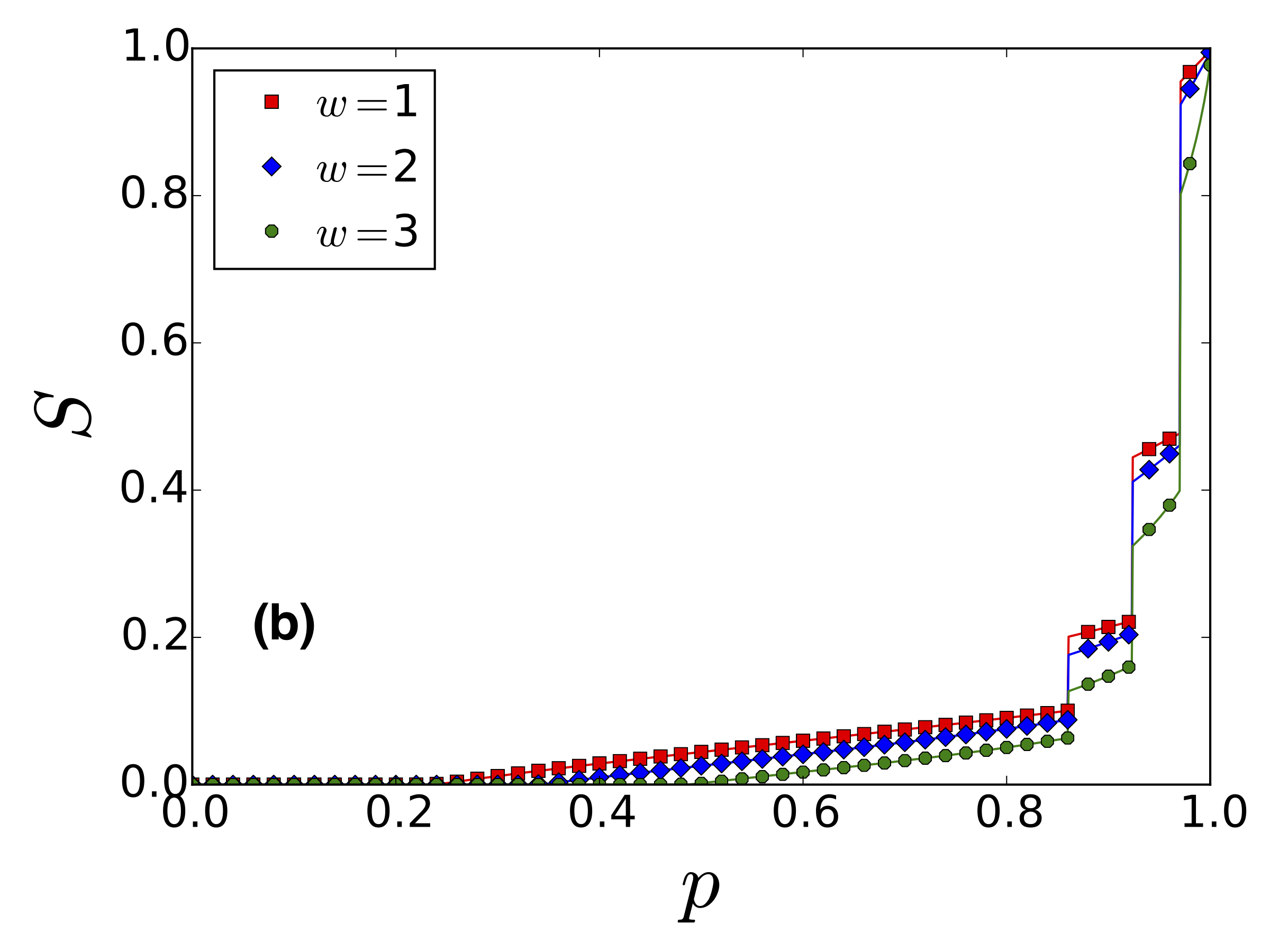} 
		\label{fig:RR_varyw}
	}	
	\caption{A random regular network of networks where each network depends on $w$ other networks such that they form loops.  We vary both \textbf{(a)} $q$ the level of interdependence between the networks (with $w=1$) and \textbf{(b)} $w$ the number of networks each network depends on (with $q=0.3$). Symbols are simulations averaged over 10 realizations on networks with $N=10^6$ nodes and lines are theory from Eq.~\eqref{eq:RR-NON}} 
	\label{fig:RR-NON}
\end{figure}

In this case, the effects of the dependencies result in a  reduction of the size of the giant component by a factor of $\left(1-q+qm_jP_\infty\right)^{w}$ \cite{gao-prl2011,gao-general-net}. Combining this with Eq.~\eqref{eq:pinf-single-layer} yields,
\begin{multline}
	P_\infty=p_{{cutoff}_{j-1}}\Bigg[\frac{1}{m_j}e^{-k_j}(1-r_j)\left(1-e^{-\left(\sum_{i=j+1}^n k_i\right) m_j P_\infty}\right)\\+r_j\left(1-e^{-\left(\sum_{i=j}^l k_i\right) m_j P_\infty}\right)\Bigg]\left(1-q+qm_jP_\infty\right)^{w}, \\ \qquad p_{{cutoff}_{j-1}}>p>p_{{cutoff}_j}. \label{eq:RR-NON}
\end{multline} 
For the RR NON the final transition will be continuous (depending on the value of $q$) as for an RR NON formed of \er networks \cite{gao-general-net}. We observe excellent agreement between the theory of Eq.~\eqref{eq:RR-NON} and simulations in Fig.~\ref{fig:RR-NON} including the prediction regarding the nature of the transitions.

\section{Discussion}
In this work, we have studied the robustness of networks and networks of interdependent networks with a hierarchical structure. This structure is very common for many infrastructure networks, biological networks and others. We have found analytical solutions and confirmed these solutions through simulations for isolated hierarchical networks and two different structures of interdependent hierarchical networks. The resilience of the network depends on the number of communities at each level of the hierarchy, the degree at each level of the hierarchy, the fraction of nodes removed, and also the parameters governing the interdependence (if present).

Our results show that hierarchical networks can undergo multiple abrupt transitions depending on the above parameters and that these transitions represent the separation of the network at different levels of the hierarchy. These results have potential applications in optimizing the resilience of networks in infrastructure and other fields.

We acknowledge the Israel Science Foundation, ONR, the Israel Ministry of Science and Technology (MOST) with the Italy Ministry of Foreign Affairs, BSF-NSF, MOST with the Japan Science and Technology Agency, the BIU Center for Research in Applied Cryptography and Cyber Security, and DTRA (Grant no. HDTRA-1-10-1- 0014) for financial support.
\bibliography{paper}

\end{document}